\documentclass[
10pt, twocolumn, prl   
]{revtex4}
\usepackage{graphicx, hyperref, color, bm, amsmath, amssymb, lipsum}   
\newcommand{\cf}[0]{{\mathcal{C}}} 
\begin{document}
\title{Flux mobility delocalization in the Kitaev spin ladder} 
\author{Alexandros Metavitsiadis}\email{a.metavitsiadis@tu-bs.de}
\author{Wolfram Brenig}\email{w.brenig@tu-bs.de} 
\affiliation{Institute for Theoretical Physics, Technical University Braunschweig,
D-38106 Braunschweig, Germany} 
\date{\today}

\begin{abstract}
We study the Kitaev spin-$1/2$ ladder, a model which exhibits self-localization 
due to fractionalization caused by exchange frustration. When a weak magnetic 
field is applied, the model is described by an  effective fermionic 
Hamiltonian, with an additional time reversal symmetry breaking term. We show that 
this term alone is not capable of delocalizing the system but flux mobility is a 
prerequisite. For magnetic fields larger but comparable to the flux gap, fluxes 
become mobile and drive the system into a delocalized regime, featuring finite dc 
transport coefficients. 
Our findings are based on numerical techniques,  exact 
diagonalization and dynamical quantum typicality, from which,  we present results for the 
specific heat, the dynamical energy current  correlation function, as well as the 
inverse  participation ratio, contrasting the spin against the fermion 
representation. Implications of our results for two-dimensional extensions of 
the model will be speculated on. 
\end{abstract}
\maketitle
\paragraph{\sc \P Introduction.}
Quantum spin liquids (QSL) are intriguing states of strongly correlated and 
highly  entangled magnetic moments lacking spontaneous symmetry breaking  
and finite local order parameters down to zero temperature $T=0K$
\cite{0034-4885-80-1-016502,RevModPhys.89.025003, 
doi:10.1146/annurev-conmatphys-031218-013401}. Instead,  
they feature topological order parameters and fractional excitations.  One renown example of a 
$\mathbb{Z}_2$ QSL is the exactly solvable, 
two-dimensional (2D),  Kitaev, spin-$1/2$ model (KSM), on a Honeycomb lattice 
\cite{Kitaev20062, 033117-053934}. 
The spins that reside on the vertices of the 
Honeycomb lattice exhibit frustrating compass interactions and as a result 
fractionalize into fermions and $\mathbb{Z}_2$ gauge fluxes. Hence the total 
Hilbert space is fragmented in subspaces with reduced or absent translation symmetry. 
While the ground state resides in a uniform flux sector, fluxes 
can be thermally excited, thus becoming a temperature activated binary disorder
for the fermions to scatter off.  

Besides the original 2D-KSM, variants of it with  different spin 
\cite{PhysRevLett.123.037203, PhysRevB.78.115116,Rousochatzakis2018} 
or dimensionality have also been discussed in the literature 
\cite{doi:10.7566/JPSJ.89.012002, PhysRevLett.98.087204, Wu20123530, 
PhysRevB.93.214425, PhysRevB.96.041115, PhysRevB.99.205129, PhysRevB.99.224418}. 
The Kitaev ladder, a one-dimensional (1D) KSM is a very interesting model because 
the reduced dimensionality inflicts additional peculiarities upon it. 
The fermionic representation with the emergent $\mathbb{Z}_2$ gauge field still 
holds but in 1D the scattering off of fermions on disorder leads to localization 
\cite{PhysRevB.96.041115}. Thus in the 1D-KSM,  single particle states are 
Anderson localized \cite{PhysRev.109.1492}  effectively leading to  many body 
localization (MBL) \cite{doi:10.1146/annurev-conmatphys-031214-014726,ALET2018498}. 
The paradigm of localization in the absence of external disorder goes back to 
two-constituent systems (light-heavy particles) \cite{JETP60} and has currently 
resurfaced in fracton phases of matter 
\cite{doi:10.1146/annurev-conmatphys-031218-013604} with numerous applications on 
lattice gauge models and more \cite{PhysRevB.90.165137, PhysRevB.91.184202, 
PhysRevLett.117.240601, PhysRevLett.118.266601, PhysRevB.96.035153, 
PhysRevB.97.104307, PhysRevLett.123.130402, PhysRevLett.120.030601, 
PhysRevB.99.054403, PhysRevX.10.011047}. 

While transition metal compounds with a Kramers doublet 
due to strong spin-orbit coupling  are good candidates for 
realizing the KSM \cite{PhysRevLett.102.017205, PhysRevLett.105.027204, 
takagi2019kitaev}, a proximate Kitaev-QSL is the closest that has been reported so far 
\cite{PhysRevB.90.041112, RevModPhys.87.1, NatureMaterials, 
PhysRevLett.108.127204, Nishimoto2016,Yadav2016,Kitagawa2018, Wulferding2020}. 
A valuable alternative for realizing the KSM might occur  
in cold atom experiments where compass interactions can be engineered 
\cite{PhysRevLett.91.090402}. Furthermore, optical lattices  are also 
advancing the experimental study of non-ergodic systems exhibiting 
MBL \cite{Schreiber842}. Remarkably, the demonstration for realizing lattice 
gauge model with $\mathbb{Z}_2$ gauge fields coupled to 1D fermions has 
recently been reported \cite{Barbieroeaav7444}. Thus, all three fascinating 
fields of Kitaev-QSL, MBL, and lattice gauge models that will be discussed 
in this work, share the prospect of experimental materialization.   

Here, we present results on the 1D-KSM including a uniform external 
magnetic field. First, using the specific heat we show that fluxes have 
a clear imprint to the specific heat. 
At weak magnetic fields, the effective fermionic representation still 
holds, with the magnetic field accounting for an additional next 
nearest neighbor (NNN), time reversal symmetry (TRS) breaking term. 
Violating time invariance in the context of Anderson 
localization could lead to delocalization due to the avoiding of 
multiple scattering events and thus reducing interference effects 
\cite{economou2006green}. Our results on the inverse participation 
ratio (IPR) as well as transport coefficients exclude this scenario.  
For larger magnetic fields, we are able to detect a delocalization transition,  
diagnosed  by finite dc transport coefficients. This is, however, attributed 
to different physics, namely, to the mobility of the fluxes. 

Our work is also of great interest for thermal measurements 
in proximate Kitaev-QSL materials, like $\alpha$-$\mathrm{RuCl_3}$. 
Despite its weak nature, the TRS breaking term, generated by the magnetic field ,  
has been allegedly reported to give rise to a quantized 
thermal Hall effect in $\alpha-\mathrm{RuCl_3}$ \cite{Kasahara2018}, 
as originally predicted for the pure 2D-KSM \cite{Kitaev20062}. 
However, the existence and the nature of it are still under investigation  
\cite{PhysRevLett.120.217205, PhysRevB.99.085136, 
yokoi2020halfinteger, PhysRevX.8.031032, PhysRevLett.121.147201}. 
Longitudinal thermal transport is also very important for the 
understanding of the excitations in these systems 
\cite{PhysRevLett.120.067202,  PhysRevLett.120.117204, 
PhysRevB.95.241112, PhysRevLett.118.187203}.  From our analysis 
we can speculate on the longitudinal thermal transport of the 2D model. 
The hallmark of the spins' fractionalization on the transport 
properties is a low frequency depletion in the spectrum of 
the dynamical energy correlations. While the limiting 
dc behavior differs for the 1D- and the 2D-KSM, the low frequency cut is 
a common attribute of both models \cite{PhysRevB.96.041115, 
PhysRevLett.119.127204, PhysRevB.96.205121, PhysRevB.99.075141}.  
Our analysis here shows that magnetic fields larger but comparable
to the flux gap make fluxes mobile, the low frequency spectrum 
depletion is filled in, and consequently the 
dc thermal conductivity is increased. After this process 
is completed, we do not expect significant changes 
in the dc thermal conductivity for further increasing  the magnetic field.   
\begin{figure}
\begin{center}
\includegraphics[width=0.95\columnwidth]{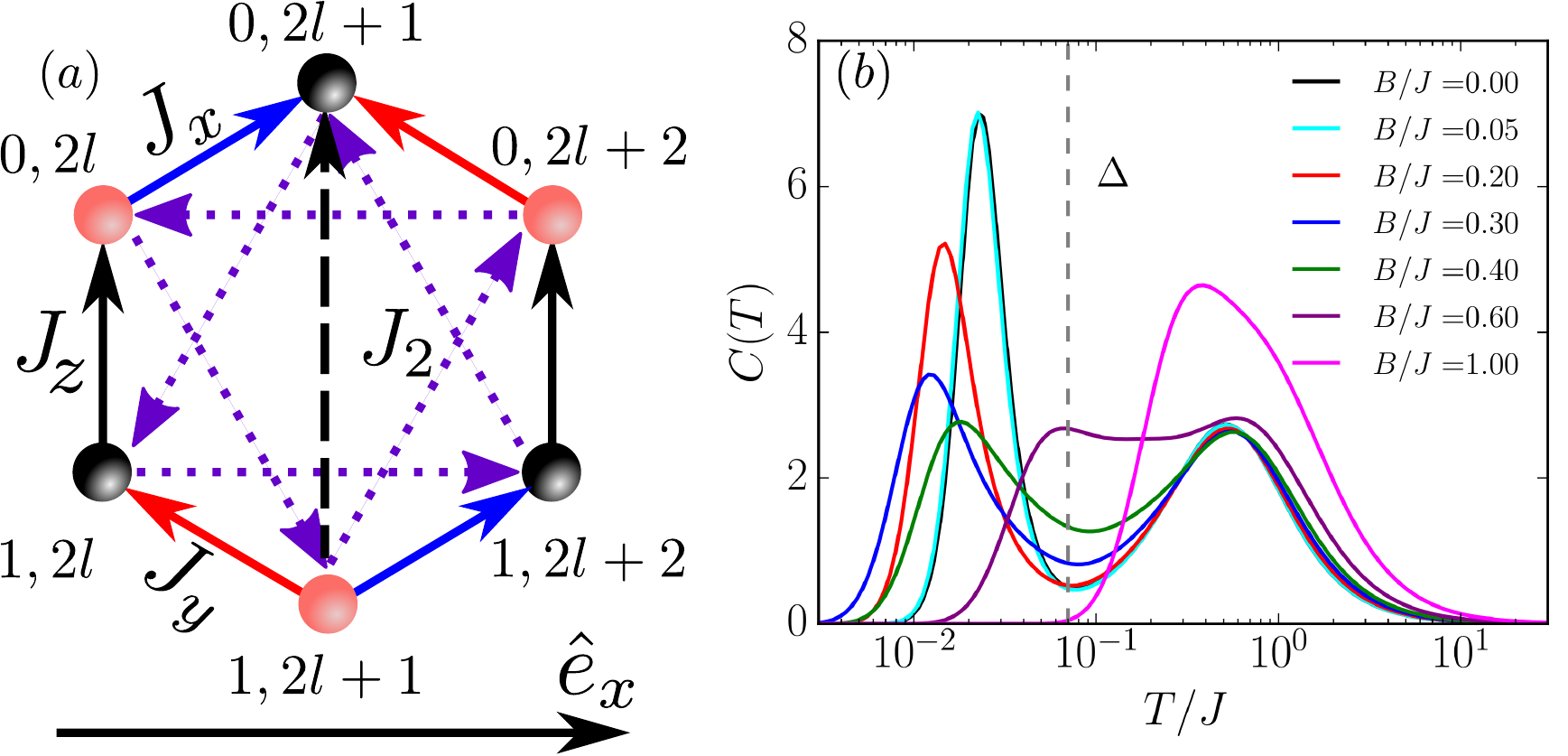}  
\caption{(a) A three rung unit cell of the spin-$1/2$ Kitaev ladder. 
The direction of the arrows indicate the order of the Majorana pairs 
in the fermionic representation. The black dashed bond on rung $2l+1$, 
connecting the 0th and the 1st chain, comes from boundary conditions 
in the rung direction. The $\hat{e}_x$ vector emphasizes the 1D 
character of the model. (b) Specific heat versus temperature 
for different values of the magnetic field $B$. The data are obtained 
via ED in the spin representation for a system of $L=8$ rungs (i.e. 16 spins).   
\label{fig:model}  
}
\end{center}
\end{figure}
\paragraph{\sc \P Model.} 
The KSM model describes bond-directional Ising interactions 
between spin $S=1/2$ operators \cite{Kitaev20062}. Its Hamiltonian in the 
presence of a magnetic field is given by  (see also  Fig.~\ref{fig:model})  
\begin{equation} \label{eq:spinHamiltonian} 
H_S = \sum_{\langle i,j \rangle } J^a_{ij} S_{i}^a S_{j}^a + g \mu_B  \mathbf{B} \cdot  
\sum_{j} \mathbf{S}_{j} 
\end{equation}
with $J^a_{ij}$ the Kitaev interactions ($a=x,y,z$), $i,j$ nearest neighbor's (NN) 
sites on the lattice, $g=1$ is the $g$-factor, $\mu_B=1$ the Bohr 
magneton, and $\mathbf{B}=(B,B,B)$ the magnetic field. We also set to 
unity the Planck and Boltzmann constants $\hbar, k_B=1$. The $J_z$-bonds 
in the middle of the hexagon arise from boundary conditions in the rung 
direction. Although the absence of these terms in Heisenberg Hamiltonians might 
give rise to new physics \cite{PhysRevB.97.214433}, here they are not 
expected to play any role. 

For $B=0$, KSM is characterized by a macroscopic number of local  conservation laws, 
the so called flux (or vison) operators and due to that it becomes analytically solvable.  
The ground state sector resides in the uniform flux sector which is 
separated from other sectors by a gap $\Delta$.   
Here, we fix the Kitaev couplings to $J_x=2J$ and $J_y=J_z=J$, where the 
ground state is gapless, and we numerically determine 
$\Delta\simeq 0.07 J$.  

At finite temperatures the fluxes become thermally excited 
and a flux proliferation process occurs for $T<\Delta$. 
This behavior can be read off from the specific heat,  
$C(T) = \left(\langle H_S^2 \rangle - \langle H_S\rangle^2\right)/ T^2$, 
which is shown in Fig.~\ref{fig:model} (b) for different values of the 
magnetic field $B$ \cite{PhysRevB.101.100408, Patel12199}. 
The results are obtained from exact diagonalization for  
an $L=8$ rung system.  For $B=0$, it exhibits the characteristic 
two-peak structure of Kitaev systems \cite{PhysRevLett.113.197205, 
PhysRevB.92.115122,  PhysRevB.96.125124, PhysRevB.96.205121, 
PhysRevB.99.094415}. The low-temperature peak is associated with the 
flux proliferation, where the system gets flooded with flux excitations. 
The action with the Zeeman term  creates an effective hopping term 
for the visons, making them effectively mobile \cite{PhysRevB.84.155121}.  
For $B<\Delta$, $C(T)$ remains practically unaffected indicating that 
the picture of the fluxes still holds. Intermediate magnetic 
fields reduce the height of the low-$T$ peak, which initially moves 
towards lower temperatures, characterizing a regime where 
visons are still present albeit  mobile.  For stronger $B$'s, the low temperature 
peak shifts to higher temperatures until it disappears, 
illustrating the absence of any trace of the fluxes.  

Treating the magnetic field perturbatively for $B<\Delta$ 
enables a fermionic 
representation where spin operators 
are mapped into two  species of Majorana fermions $c$ and 
$\bar c$ \cite{Kitaev20062, PhysRevB.86.085145}, with  
$\{c_i,c_j\} = 2\delta_{ij} = \{\bar{c}_i,\bar{c}_j\}$ and 
$\{c_{i},\bar{c}_{j}\} = 0$.  One of the two species, say $c$, is itinerant, 
while the other pair up along the $z$-bond direction,  they commute with the 
Hamiltonian, and they become static. We denote these local conservation laws 
with $\eta_j^z =\pm 1$ while we also introduce $\eta_{j}^{x}= \eta_{j}^{y}=1$ 
to unify the notation. The  ground state occurs for $\eta_j^z=1$ while for
$T \gg \Delta $ $\eta_j^z$ is completely disordered, 
$\langle \eta^z \rangle =0$. The magnetic field accounts for a TRS-breaking, 
NNN-interaction term in the fermionic representation, i.e., $H_S \approx H_F$ with 
\begin{equation} \label{eq:fermionicHamiltonian}
H_F  = -\frac{i}{4}  
\sum_{\langle i,j \rangle}  
J^a_{ij} \eta_{i}^a c_{i} c_{j} -i\frac{J_2}{8}  
\sum_{\langle \langle i,j \rangle \rangle}
\tilde{\eta}_{ij} c_{i} c_{j}~, 
\end{equation}
and $J_2 \sim  \frac{B^3}{\Delta^2} $ \cite{Kitaev20062, PhysRevB.86.085145, noteJ2}.
The double brackets in the second term denote summation over 
NNN sites and the order of the majorana pairs can be read 
off from Fig.~\ref{fig:model}(a).
For $\tilde{\eta}$  holds: 
$\tilde{\eta}_{ij}=1$ for intrachain bonds  or 
$\tilde{\eta}_{ij}= \eta_{i}^z + \eta_{j}^z$ for interchain bonds.
In terms of Dirac fermions \cite{PhysRevLett.98.087204, Chen_2008, 
PhysRevB.79.214440, Mandal_2012, RevModPhys.87.1}, 
$H_F$ becomes a superconducting Hamiltonian 
on a two-site unit cell chain of length $L$, in the presence 
of an onsite $\mathbb{Z}_2$ gauge field $\sim \eta_j^z$ as well as 
bond disorder terms $\sim \tilde{\eta}_{ij}$. 
\begin{figure}
\begin{center}
\includegraphics[width=\columnwidth]{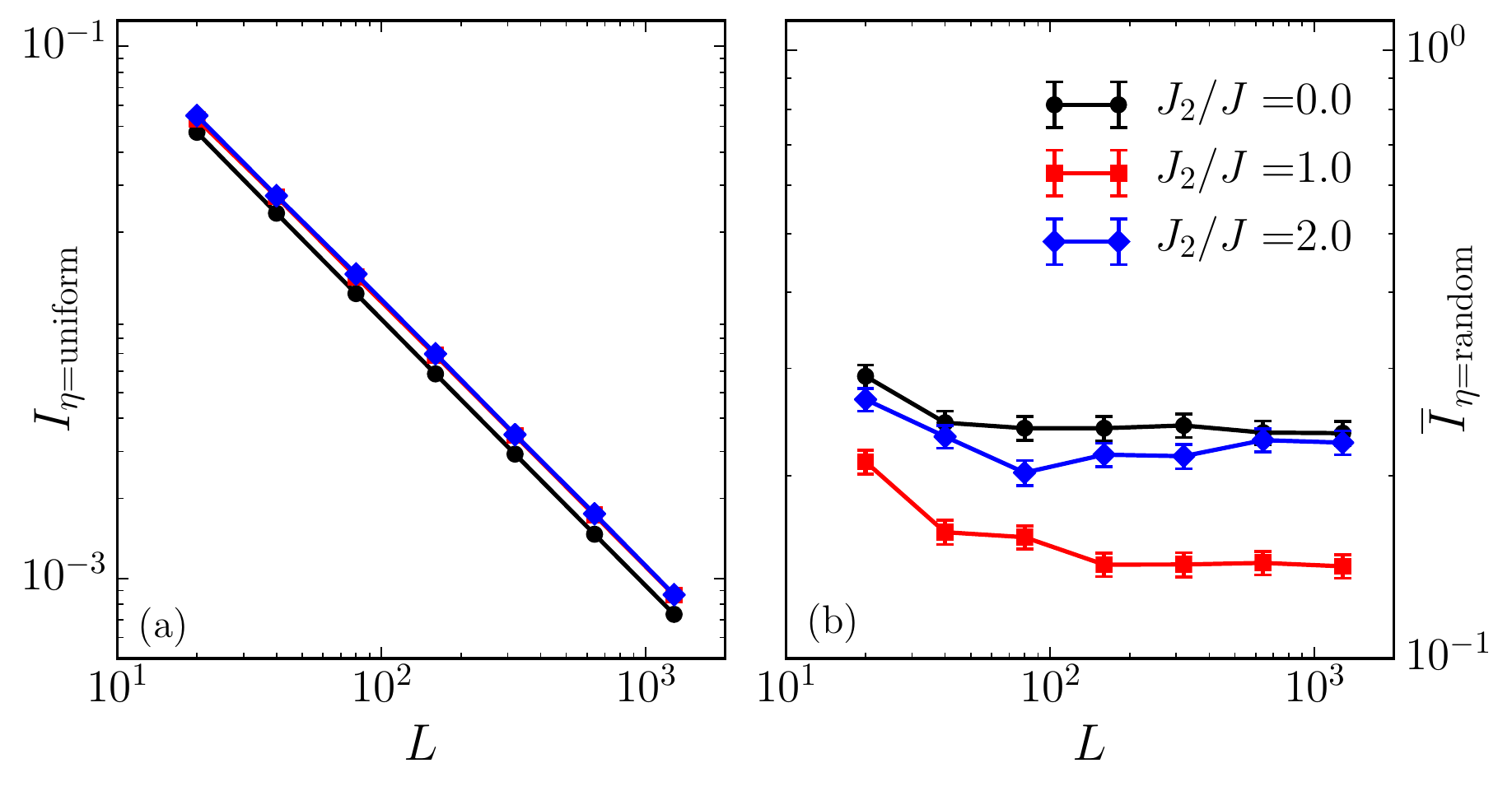}  
\caption{System size scaling of the inverse participation ratio of the fermionic model, 
Eqs.~\eqref{eq:fermionicHamiltonian} and \eqref{eq:IPR}, in a log-log scale 
for different values of the $J_2$ coupling. (a) Uniform gauge configuration. 
(b) Random average over $R=1000$ maximally disordered states, $\langle \eta^z \rangle = 0$, 
and  $3\sigma$ depicted as error bars.  
\label{fig:IPR} 
}
\end{center}
\end{figure}
\paragraph{\sc \P Inverse participation ratio.}

The first quantity that we look at in order to detect localization 
is the inverse participation ratio (IPR), which is given by the sum 
over the lattice sites of the squared probabilities of the 
wave-functions $\psi$ \cite{RevModPhys.80.1355}. For a given $\eta$-configuration,  
we denote the average IPR  with $I_\eta$, while for disordered sectors, we  average 
over $R$ gauge configurations to obtain the moments $\mathcal{I}_p$, viz., 
\begin{equation} \label{eq:IPR}
I_{\eta} = \frac{1}{L} \sum_{m=1}^L  \sum_{l=1}^{L} |\psi_m^{\eta}(l)|^4,\quad 
\mathcal{I}_p = \frac{1}{R} \sum_{r=1}^R \left(I_{\eta_r}\right)^p  ~. 
\end{equation} 
From these definitions, the mean IPR is given by $\overline{I} = \mathcal{I}_1$ 
while the fluctuations around this mean can be quantified via the 
standard deviation $\sigma = \sqrt{\delta \overline{I}/R}$, where 
$\delta \overline{I}=  \mathcal{I}_2 - (\mathcal{I}_1)^2$. 

Assuming that all states of a system are localized  
$ \psi_m^{\eta}(l) \sim \delta_{lm}$, the IPR is expected to scale as 
$\overline{I}(L)  \sim \text{const.}$, while for extended states
$ \psi^{\eta}_{m}(l) \sim 1/\sqrt{L}$,   $\overline{I}(L) \sim 1/L$.    
In Fig.~\ref{fig:IPR}(a), we plot in a log-log scale 
the IPR of the uniform gauge sector for different values of the $J_2$ coupling 
versus the system size, which reveals a $\sim 1/L$ scaling. On the contrary, for  
the same values of $J_2$, a random averaging over $R=1000$ sectors with 
$\langle \eta^z \rangle=0$ reveal the opposite behavior, namely 
$\overline{I}(L)  \sim \text{const.}$. 
The difference between ``clean'' and ``dirty'' sectors is striking, and elucidates 
the localization character of the disordered states for any $J_2$.
The initial drop of the IPR in Fig.~\ref{fig:IPR}(b) can be attributed to a 
comparable localization length $\xi$ with the system size. Moreover, from 
this behavior, it is hard to conclude a large sensitivity of $\xi$ to $J_2$.  

\paragraph{\sc \P Energy transport.} 
\begin{figure}
\begin{center}
\includegraphics[width=\columnwidth]{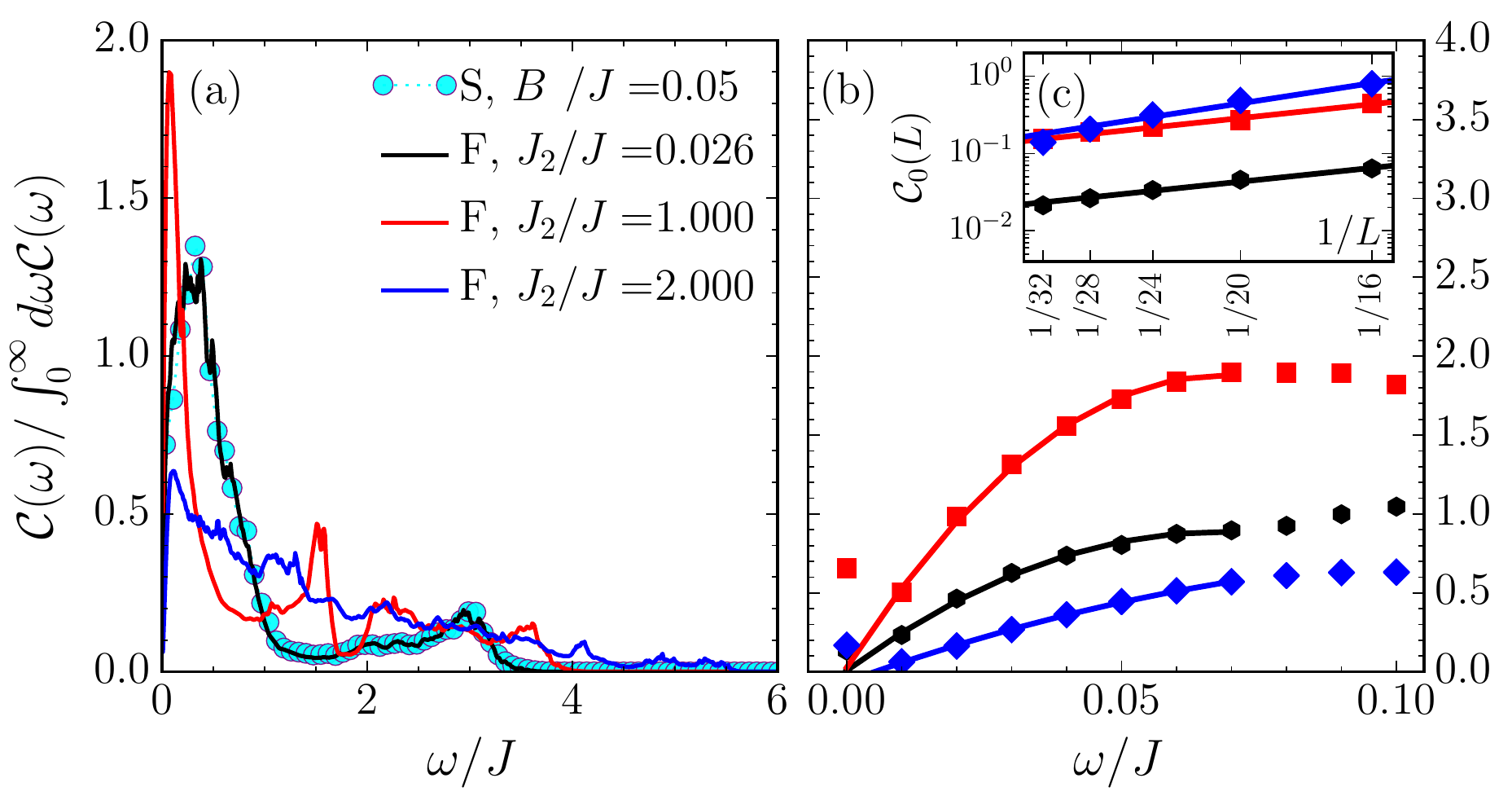}  
\caption{(a) Frequency dependence of the energy current correlation 
function in the fermionic representation, Eq.~\eqref{eq:fermionicHamiltonian}, 
for  $J_2/J=0.026,1,2$ and $L=32$. For comparison, $\cf(\omega)$ evaluated 
with $H_S$, for $B=0.05J$, and $L=8$ is also shown marked with cyan circles 
connected by a dotted line. (b) A zoom at low frequencies of (a). The lines 
depict second order polynomials, fitted in the range $0<\omega/J < 0.07$ to 
extract the dc limit. (c) Finite size scaling of the Drude weight $\cf_0$, 
in a semi-log plot. The lines are exponential fits. 
}    
\label{fig:cFermionic} 
\end{center}
\end{figure}
Next we study the dynamical transport properties of $H_F$. For that, 
we employ the energy current dynamical auto-correlation function, which 
has the advantage to be diagonal in the gauge fields, and it is also 
directly related to the experimentally measurable thermal conductivity, 
\begin{equation}
\cf(t) = \frac{1}{L} \left<  j^\epsilon(t) j^\epsilon \right > , \quad 
\cf(\omega) =  \int dt e^{i\omega t} \cf(t) ~. 
\end{equation}
Here, $j^\epsilon$ is the energy current operator, the exact expression 
of which is acquired via the time derivative of the polarization operator  
$j^\epsilon=\frac{\partial P}{\partial t}$ , $P = \sum_l r_l h_l$, 
with $h_l$ being a local energy density \cite{noteUnitCell} 
and $r_l$ its corresponding coordinate. The angled brackets denote 
a thermal expectation value, which here is restricted to infinite 
temperature. For the discussion of localization 
we are interested in the low-$\omega$ properties of $\cf(\omega)$, 
and mainly its static part, which comprises two contributions: 
(i) the Drude weight arising from the non-vanishing due to degeneracies 
part of the correlation function at longer times, 
$ 2\pi \cf_0 = \lim_{t\rightarrow \infty} \cf(t)$; (ii) 
the dc limit of the regular part 
$\cf_{dc} = \lim_{\omega \rightarrow 0} \cf(\omega)$. The former 
indicates ballistic while the latter dissipative transport and 
if both of them vanish, the system is an insulator.  

In Fig.~\ref{fig:cFermionic}(a), we present results for $\cf(\omega)$, 
acquired via ED in the fermionic representation, for different values of 
the NNN interaction and $L=32$, corresponding to a Hilbert space dimension 
of $2^{64}$. Due to the different energy scales, we normalize the curves 
to a unit integral. In the fermionic representation, the quadratic form 
of $H_F$ yields two types of contributions in $\cf$, 
``quasiparticle'' or ``pair-breaking''. 
These can be discerned in the curve for $J_2=0.026J$, corresponding to 
$B=0.05J<\Delta$. First, the maximum around $\omega \approx 0.4 J$, is 
attributed to the quasiparticle part of the correlation function. 
The sharp decrease of $\cf(\omega)$ at lower frequencies,  
$\omega \lesssim 0.2J$, better highlighted in Fig.~\ref{fig:cFermionic}(b), 
is inevitable due to the localization of the single particle states. 
Exactly the same behavior is recovered in the spin representation 
from the many-body Hamiltonian $H_S$, also plotted in 
Fig.~\ref{fig:cFermionic}(a), for $B/J=0.05$, and $L=8$. 
Second, the broader and of lower intensity hump, centered around 
$\omega \approx 3J$, corresponds to the pair-breaking type of contributions.  
As $J_2$ is further increased,  the gap between the quasiparticle 
and the pair-breaking contributions is filled, however, as better 
seen in Fig.~\ref{fig:cFermionic}(b), the pseudo-gap at low frequencies 
does not close. In Fig.~\ref{fig:cFermionic}(b), we highlight 
the low frequency behavior of $\cf(\omega)$ for the fermionic spectra 
plotted in panel (a). The lines connecting the points 
are second order polynomial fits in the range $0< \omega/J < 0.07$, 
and extrapolate to tiny or even negative values at $\omega = 0 $ 
\cite{noteFiniteDC}. A small Drude weight $\cf_0$ becomes also visible 
in Fig.~\ref{fig:cFermionic}(b), which finite size scaling behavior 
is plotted in Fig.~\ref{fig:cFermionic}(c) for systems $L=16-32$. 
The lines are exponential fits to the data and imply an 
exponential decay of $C_0$, namely, $C_0\rightarrow0$ 
for $L\rightarrow \infty$. Thus, it can be inferred that both  
$\cf_0$ and $\cf_{dc}$ vanish in the thermodynamic limit. 

As one of our prime results, we summarize the preceding 
by stating, that  both the IPR and $\cf(\omega)$ evidence,  
that a delocalization of the system cannot be captured within the 
fermionic representation despite the TRS-breaking nature 
of the NNN interaction induced by the magnetic field. 
\begin{figure}
\begin{center}
\includegraphics[width=\columnwidth]{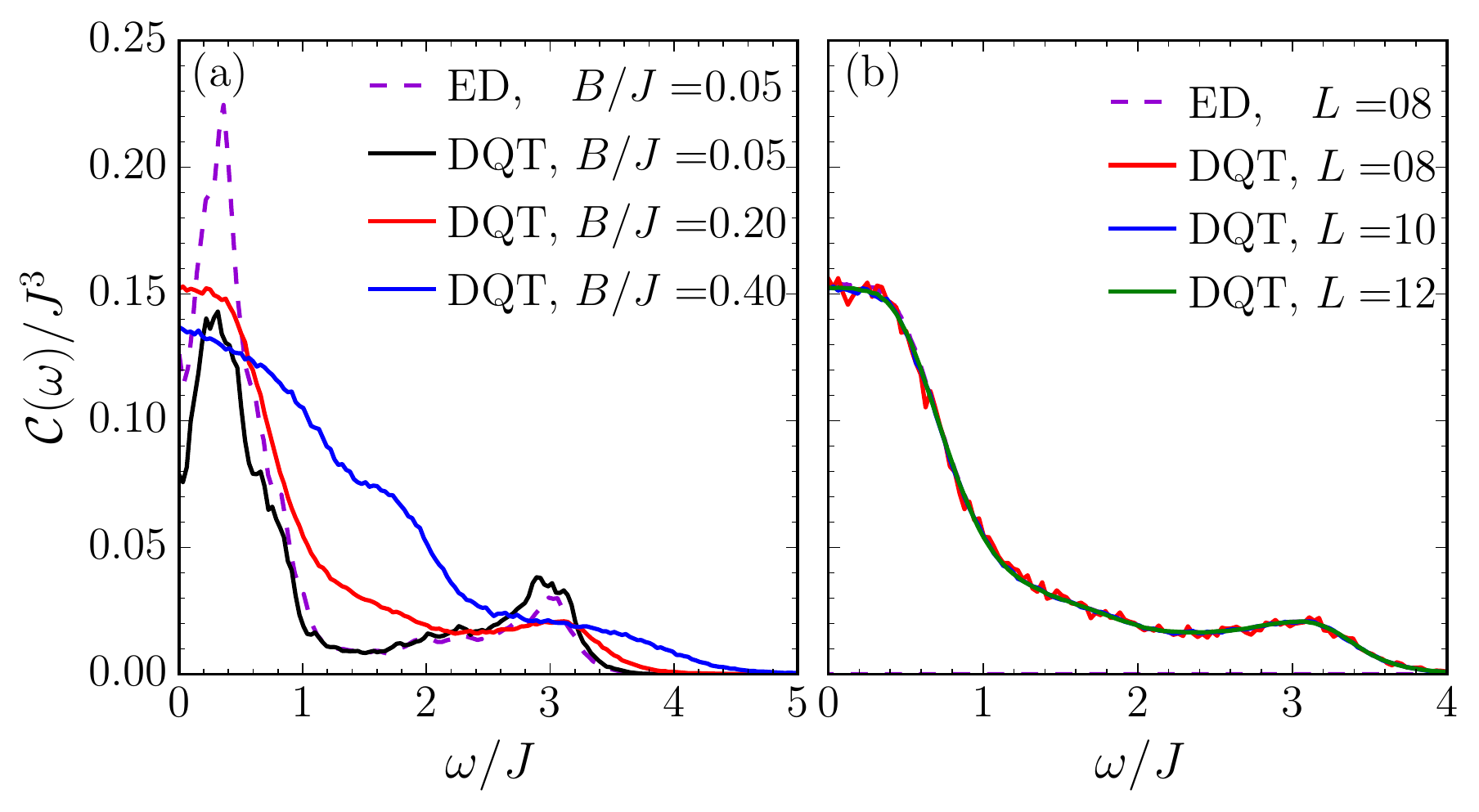}  
\caption{(a) Frequency dependence of the energy current correlation 
function in the spin representation, Eq.~\eqref{eq:spinHamiltonian} for 
different values of the magnetic field, with $L=8$ (ED) and $L=12$ (DQT). 
(b) Finite size scaling of $\cf(\omega)$ for $B=0.2J$.
}    
\label{fig:cSpin} 
\end{center}
\end{figure}

\paragraph{\sc \P Spin representation.} 
Next, we contrast the previous findings to $\cf(\omega)$ obtained within the 
$H_S$ framework, where the magnetic field is taken fully into account, 
Eq.~\eqref{eq:spinHamiltonian}. 
To improve upon the available system sizes, facing the complete 
many body Hilbert space, and in addition to ED, we also employ  
Dynamical Quantum Typicality (DQT). In DQT a 
thermal mean value is approximated by an expectation value obtained from a 
single pure random state $|\psi\rangle$, drawn from a distribution that is 
invariant under all unitary
transformations in Hilbert space (Haar measure), which leads to an 
exponential error decrease with $L$ \cite{PhysRevLett.112.120601}. 
The real part of correlation function is then evaluated via 
$\cf'(t) \approx \mathrm{Re} \frac{\langle\psi| j(t) j
|\psi \rangle}{L\langle\psi |\psi \rangle}$ by solving a standard
differential equation problem for the time evolution. 
The time evolution is performed with a $J\delta t=0.01$ step [corresponding to 
an accuracy of the order of $O(10^{-8})$ in the fourth order Runge-Kutta 
algorithm], while we integrate for times up to $Jt_{max}=100$. 

The results for $\cf(\omega)$ for different values of $B$ are presented in 
Fig.~\ref{fig:cSpin}(a). First, comparing DQT ($L=12$) and ED ($L=8$) at 
weak magnetic fields reveals a discrepancy between the two methods. 
This is due to the long time oscillations of the 
Kitaev terms, causing $\cf(t)$ to oscillate even at the longest times 
kept here. These discrepancies disappear at higher magnetic fields 
where $\cf(t_{max})=0$ and do not invalidate any of our conclusions, 
see also Fig.~\ref{fig:cSpin}(b). 
As the magnetic field exceeds the gap, we observe 
a very rapid filling of the low frequency depletion. This can be interpreted 
as a large weakening of the fluxes' scattering strength once they become mobile. 
Already at $B=0.2J$, the low frequency depletion disappears giving $C(\omega)$ a 
more Drude-like shape, although the higher frequency pair-breaking structure  
can still be observed. For even higher values of the magnetic field 
higher and lower frequencies are smoothly connected, while the dc limit 
shows only a weak dependence on $B>\Delta$. Thus one can argue, that $C_{dc}$,  
or equivalently the experimentally measurable thermal conductivity $\kappa_{dc}$, 
would exhibit a strong increase at $B \gtrsim \Delta$ followed by a weak 
variation  as $B$ is further increased.  
Lastly in Fig.~\ref{fig:cSpin}(b), we present the finite size scaling behavior of 
$\cf(\omega)$, also comparing ED and DQT. We find, that there are practically 
no finite size effects at moderate magnetic fields in the data of the spin representation. 

\paragraph{\sc \P Discussion.} 
Our main finding here is, that fractionalization 
in the 1D-KSM, leading to a thermally activated flux disorder, induces  
self-localization for $B\lesssim\Delta$ and therefore leads to  vanishing 
thermal transport coefficients. For stronger magnetic fields 
$B \gtrsim \Delta$, flux mobility and many-body interactions 
fill the low frequency depletion of $\cf(\omega \ll 1)$ leading 
to delocalization and finite transport coefficients. The 
absence of this behavior in the popular simplification of the 1D-KSM, 
which treats magnetic fields only perturbatively, raises questions 
on the applicability of the latter model to describe finite-field transport.  

Let us now speculate on the application of our results to the 2D-KSM. 
The characteristic low frequency depletion, attributed to the scattering 
of fermions on the gauge field \cite{PhysRevB.93.214425, PhysRevB.96.041115}, 
is also a characteristic of the 2D-KSM \cite{PhysRevLett.119.127204, 
PhysRevB.96.205121,PhysRevB.99.075141}.
An essential difference in the absence of magnetic field between the 
1D- and 2D-KSM is that in the latter, the pseudo-gap closes 
in the thermodynamic limit restoring dc transport. 
However, the mechanism of filling the low frequency depletion in 
the $\cf(\omega)$ spectra due to the flux mobility will be also 
present in 2D. Taking into account that the flux gap of 
the 1D and the 2D systems are almost equal, 
$\Delta_{2D} \approx \Delta=0.07J$, and that typical values of Kitaev 
exchange are $J \approx 70K$, we expect $\Delta \approx 5K$. 
Therefore, a system with purely  Kitaev interactions would exhibit a notable 
increase in the dc transport coefficients for magnetic fields around $B\gtrsim 7T$. 
\section*{Acknowledgments}  
A.M. acknowledges useful discussions on the IPR with Peter G.~Silvestrov. 
Work of W.B. has been supported in part by the DFG through Project A02 
of SFB 1143 (Project-Id 247310070), by Nds.~QUANOMET, and by the 
National Science Foundation under Grant No.~NSF PHY-1748958. 
W.B.~also acknowledges the kind hospitality of the PSM, Dresden. 
%
%
%

\begin{thebibliography}{75}
\bibitem[{not({\natexlab{a}})}]{noteJ2}
\bibinfo{note}{Although $J_2 \sim O(0.01 J)$ for $H_F$ to be valid, here we
  treat it as a free parameter to be able to detect a possible delocalization
  due to the TRS term.}
\bibitem[{not({\natexlab{b}})}]{noteUnitCell}
\bibinfo{note}{The unit cell chosen here is a linear combination of the one
  shown in Fig.~\ref{fig:model} with the last $z$-bond removed and the same one
  only shifted by one-rung, see also Refs.~\cite{PhysRevB.96.041115,
  PhysRevB.99.205129}. We have also tested other choices, which give no
  qualitative difference but at most some quantitative discrepancies at high
  frequencies.}
\bibitem[{not({\natexlab{c}})}]{noteFiniteDC}
\bibinfo{note}{Even the tiny finite values can be shown that go to zero as
  $L\rightarrow \infty$ by using an averaging over gauge configurations which
  allows to reach much larger system sizes \cite{PhysRevB.96.041115}.}
\end{thebibliography}

%
%
%
\end{document}